# Quantifying Retrospective Human Responsibility in Intelligent Systems

Nir Douer and Joachim Meyer, *Senior Member, IEEE*

*Abstract* —**Intelligent systems have become a major part of our lives. Human responsibility for outcomes becomes unclear in the interaction with these systems, as parts of information acquisition, decision-making, and action implementation may be carried out jointly by humans and systems. Determining human causal responsibility with intelligent systems is particularly important in events that end with adverse outcomes. We developed three measures of retrospective human responsibility when using intelligent systems. The first measure concerns repetitive human interactions with a system. Using information theory, it quantifies the average human's unique contribution to the outcomes of past events. The second and third measures concern human causal responsibility in a single past interaction with an intelligent system. They quantify, respectively, the unique human contribution in forming the information used for decision-making and the reasonability of the actions that the human carried out. The results show that human retrospective responsibility depends on the combined effects of system design and its reliability, the human's role and authority, and probabilistic factors related to the system and the environment. The new responsibility measures can serve to investigate and analyze past events involving intelligent systems. They may aid the judgment of human responsibility and ethical and legal discussions, providing a novel quantitative perspective.**

*Note to Practitioners* **— We developed a theoretical model and quantitative measures that can aid in assessing the retrospective human causal responsibility in the interaction with intelligent systems. Practitioners can apply responsibility measures to estimate different aspects of user causal responsibility in past interactions with intelligent systems. One must assume stationarity and ergodicity to apply the measures (which are based on Information Theory) to real-world systems. Nevertheless, one can often limit the computation to periods that satisfy stationarity, which facilitates computation. When this is not possible, the computed responsibility measures should be treated with caution and be combined with sensitivity analyses of how changes in the input probabilities and assumptions affect the responsibility values, which will often supply valuable insights.**

N. Douer and J. Meyer are with the Department of Industrial Engineering at Tel Aviv University, Ramat Aviv, Tel Aviv 69978, Israel (e-mails: nirdouer@mail.tau.ac.il, jmeyer@tau.ac.il).

## I. INTRODUCTION AND RELATED LITERATURE

In the autumn of 2011, the Swiss bank UBS reported losses of over \$2 billion due to unauthorized trading by an individual trader, a director of the bank's Global Synthetic Equities Trading Team in London. Although the trader was mainly at fault, a later inquiry revealed that UBS failed to act upon an earlier warning issued by their internal risk and operations systems. The failure to respond to the warning led to the resignation of the bank's CEO and the co-heads of the Global Equities team [1].

However, was this resignation justified? Should people always be expected or forced to resign if they did not follow a warning from an intelligent decision support system (DSS) and a negative outcome occurred?

In fact, many warning systems are problematic. For example, the abundance of "non-actionable" alarms (alarms that do not require a staff response) in intensive care units and the resulting "alarm fatigue" is a major problem in intensive care units and other medical settings [2]–[4]. Similarly, despite technological advances in banking, current anti-fraud systems produce too many false positives, hampering the efficiency of fraud detection [5]. Hence, bank employees are the final ones to decide whether a transaction the anti-fraud systems flagged is an attempted fraud. The expected outcomes of their actions affect the employees' decisions, including issues such as the fear of jeopardizing their supervisor's trust, professional status, and reputation [6]. Employees' decisions are also influenced by their inability to fully supervise advanced transaction monitoring systems that use artificial intelligence and machine learning to detect anomalous and outlying behaviors and provide real-time alerts but still yield many false positives. The substantial negative consequences associated with a wrong decision may lower employees' tendency to adhere to a fraud alert [7].

Fraud detection systems are just one example of intelligent DSSs that have become a significant part of financial markets (e.g., algorithmic trading), transportation (e.g., driving assistance systems), medical care (e.g., advanced data-based recommendation and classification systems), and many other domains, such as military, e-commerce, and entertainment. In these systems, human responsibility for the outcomes becomes unclear. For instance, what is the human responsibility when all information about an event arrives through a DSS that employs artificial intelligence to collect and analyze data from numerous sensors beyond human ability to monitor and control independently? If a human act the system indicated as necessary eventually turns out to be wrong, is the human (entirely or partly) responsible for the outcomes?

### A. Responsibility in intelligent systems

Responsibility has different aspects, such as role responsibility, causal responsibility, liability, moral responsibility, and capacity responsibility [8], [9]. In human interactions with systems, *role responsibility* is allocating specific duties to the human. However, these duties do not specify the causal relations between human actions and their influence on the outcomes, which are defined by *causal responsibility*. This paper focuses on *retrospective* causal responsibility for past events [10].

The philosophical and legal literature contains detailed discussions of retrospective causal responsibility, a major factor in how legal doctrines determine liability, punishments, and civil remedies in criminal and tort law [11]–[13]. Also, extensive literature in psychology, such as attribution theory, sees causal responsibility as an essential primary condition for attributing blame and praise [14]–[17].

For humans to have causal responsibility for outcomes when interacting with systems, they must have some control over the system and resulting outcomes [18]. However, as systems become more intelligent, the human ability to control processes diminishes, as these become opaque ("black box"), beyond the human's supervising capabilities [19]. In addition, the transition to *shared control* in advanced systems, in which humans and systems jointly make decisions and actions, further complicates the ability to determine the unique human contribution to the joint processes [20]. Hence, humans may no longer be considered entirely responsible for the outcomes of intelligent systems [21]–[23], and there is a "*responsibility gap*" in the ability to assess the actual human retrospective responsibility [24]–[26].

### B. Quantitative estimation of retrospective responsibility

In criminal law, causal responsibility for a past event must be proven with a high standard of proof, often called "beyond a reasonable doubt." A lower certainty suffices in civil matters, in which causal responsibility for an event must be proven to be only more probable than the alternative. This is the case in tort law that provides persons with remedies to compensate for breaches of legal duties owed to them by others. The general tort rule is that it is enough for the proof to tip a balance of probabilities of 50%, in which case the more probable alternative is treated as a certainty [12]. Under the general law, damages are either awarded in full or none. This all-or-nothing rule has clear limitations. First, it seems wrong to treat one alternative as certain if it has only a slightly higher probability of occurring and to neglect the contributing causality of the rejected alternative, which has a somewhat lower probability [27], [28]. In addition, there are cases in which the tort was caused by at least one of several persons, each of whom may have behaved in ways that could have led to adverse outcomes [12]. A legal approach that tries to overcome these limitations is the use of comparative responsibility (or comparative negligence), a doctrine of tort law that divides causal responsibility for the damages among different parties in proportion to their contribution to the outcomes [29], [30].

When assessing comparative responsibility, judges and juries often use expert witnesses and consultants, which often employ common statistical methods, such as descriptive statistics, causal inference, predictions of the likelihood of an event, considering earlier occurrences, and Bayesian inference [31], [32]. However, since real-world conditions are often complex and context-dependent, different experts may reach different conclusions depending on the method used, which may confuse jurists and judges [31].

Common statistical methods are also inadequate for quantifying comparative human causal responsibility in complex forms of human interactions with intelligent systems or in rare malfunction scenarios, which lack sufficient numbers of past occurrences to enable sound statistical reasoning.

Recently, we developed and tested a new model that quantifies *prospective* causal responsibility in interactions with intelligent systems (the ResQu model) [33]–[36]. Given a specific system design and human capabilities, this model uses information theory to predict the expected share of unique human contribution to the overall outcomes. Our paper aims to expand the ResQu model [33] by developing *retrospective* measures to quantify human responsibility in a *specific* past interaction, for which one knows the detailed chain of events and outcomes. Such an analysis is, for instance, critical when determining whether disciplinary actions should be taken after an adverse outcome involving an intelligent system and a human.

## II. A MODEL OF RETROSPECTIVE RESPONSIBILITY

### A. A general model of information flow

We hereby summarize the parts of the ResQu model [33] as a basis for the new analysis of retrospective responsibility.

*Environment states*: The environment includes $N$ possible states ($N{\geq}2$), with probability distribution $\{p_i\}_1^N$. Each of the $N$ states is characterized by $m$ different observable and measurable parameters $E_i$ ($i$=1...m). Different states have partly overlapping distributions of the observable parameters. Thus, there is uncertainty regarding the actual environmental state.

*Information acquisition*: The variables $Y_i$ and $X_i$ ($i$=1...m) denote, respectively, the values of $E_i$ ($i$=1...m) that the intelligent system and the human acquired. Not all the measurable parameters may be observable by the human or the intelligent system.

*Information analysis*: The variables $Y_a$ and $X_a$, denote, respectively, $N$ dimension vectors, generated by the intelligent system and the human, that assign posterior probabilities to each of the possible $N$ environment states. The system's results may serve as additional input for the human information analysis and vice versa.

*Action selection*: The variables $Y_s$ and $X_s$ denote, respectively, the action alternatives of the intelligent system and the human. We denote by $y_s$ and $x_s$, the corresponding actual selected actions. For the system, $y_s$ is exclusively defined by its algorithm once all input variables are acquired. For the human, the selected action $x_s$ is based on the results of the human information analysis and on the human utility function, which relates costs and benefits to different outcomes.

*Action implementation* We denote by $Z$ the implemented action. It depends on the actions the system and the human selected and the system design, which determines how they are combined and prioritized.

Fig. 1 presents a schematic depiction of the variables and information flow in the ResQu model.

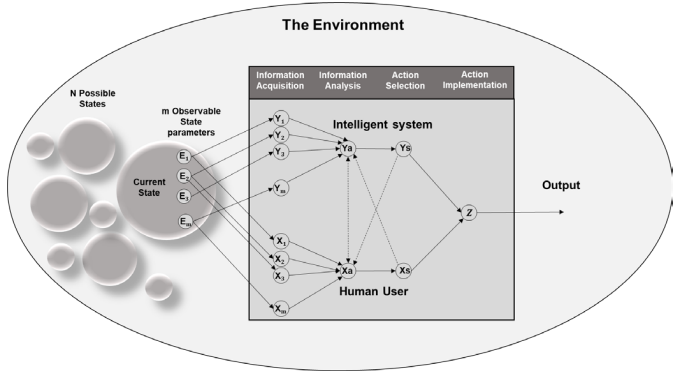

Fig. 1. General model of information flow in human interaction with an intelligent system. Dashed lines represent possible information transfer between the human and the intelligent system.

### B. Retrospective responsibility in a repetitive task

We first discuss a case in which humans repeatedly interacted with an intelligent system. In this case, one can infer the underlying distributions of the model's variables from empirical observations made over time and calculate the average human *retrospective* responsibly, *Resp(Z)*, defined as the unique human contribution to the outcomes.

$$Resp(Z) \stackrel{\text{def}}{=} \frac{H(Z/Y_1 \dots Y_m, Y_a, Y_s)}{H(Z)} \qquad (1)$$

where $H(X)$ is Shannon's entropy, which measures uncertainty related to a discrete random variable X, and H(X/Y) is the conditional entropy, which measures the uncertainty remaining about a variable $X$ when a variable $Y$ is known.

By definition, *Resp(Z)* $\in [0,1]$. It equals 1 if $H(Z/Y_1 \dots Y_m, Y_a, Y_s) = H(Z)$, which occurs when the distribution of the output variable Z, observed over many trials, is unaffected by the intelligent system variables. Thus, the system has no real contribution, and all uncertainty about $Z$ is related only to the human variables, so the human has full casual responsibility for the output. On the contrary, *Resp(Z)* equals 0 if $H(Z/Y_1 \dots Y_m, Y_a, Y_s) = 0$, which occurs when the intelligent system variables solely determine the distribution of Z. In this case, human information processing and action selection did not have any meaningful contribution, as the same outcomes would be achieved without human involvement. Values between 0 and 1 represent intermediate levels of human contribution to the overall output.

To calculate this average retrospective responsibility measure, one must either infer the underlying distributions of the model's variables from empirical observations taken over time or known properties. Since it is based on entropy, a caveat is that it is not directly applicable when human or system characteristics are not stationary or ergodic (e.g., when learning or fatigue effects lead to a change in the level and contribution of human involvement over time). Nevertheless, one can often limit the computation to periods that satisfy stationarity. For example, in laboratory experiments in which a classification system aided participants, the participants' average retrospective responsibility was computed only for data collected after the learning effect subsided [34].

### C. Retrospective responsibility for a single past event

Computation of *Resp(Z)* in (1) is not feasible in a *single* past event of human interaction with a system. In this case, only a single realization of each variable is observed, so we cannot deduce the underlying distributions required for calculating entropy and mutual information. Hence, additional retrospective responsibility measures are needed.

In real-life situations, retrospective inquiries can identify cases in which the intelligent system entirely dictated the implemented action (e.g., when a car initiated an automated emergency braking without the driver's involvement). In these cases, the system solely decided and carried out the selected action, and the human comparative contribution to the outcomes is zero.

Hence, we focus on the more complicated cases in which the implemented action $Z$ was dictated by the human action selection $X_s$. This is, for example, the case with DSSs that only present additional information for the human to consider before selecting an appropriate action.

We propose two complementary retrospective responsibility measures. The first measure quantifies the unique share of the human in providing and evaluating the information used for decision-making. It is a measure of the human retrospective unique contribution to the result of the human information analysis stage. The second measure quantifies how reasonable the specific action the human selects is given the available information to the human at the decision time.

### Retrospective responsibility in the information analysis

In the ResQu model, the variable $X_a$ denotes a probability distribution generated by the human, which assigns posterior probabilities to each possible N environment state. This distribution serves as the base for the human action selection and the human utility function, which relates costs and benefits to the predicted outcomes of each action alternative at each environmental state. We will denote by $x_a$ a single realization of $X_a$, based on specific given values of acquired information and the particular distribution of different states in the environment. The vector $x_a$ is computed using all information, known to the human at the decision moment. This includes the specific values acquired by the human, $\{x_i\}_1^m$, and, depending on the automation level, function allocation, and user interface, may also include specific system values $\{y_i\}_1^m$, $y_a$ and $y_s$. These are the values the intelligent system acquired, the results of its information analysis, and its selected action (see Figure 1).

Denote by $x_{aH}$ the posterior probability distribution of the environmental states when the human uses only human self-acquired information, $\{x_i\}_1^m$. Denote by $x_{aS}$ the posterior probability distribution of the environmental states when the human uses only information from the intelligent system. The comparative human contribution in generating $x_a$ decreases as $x_a$ approaches $x_{aS}$. When $x_a = x_{aS}$, the information the human acquired does not contribute to generating $x_a$. As the difference between $x_a$ and $x_{aS}$ increases, the information from the system has less impact on the posterior probabilities the human assigns to each environment state.

We use the Jensen-Shannon distance, $D(P,Q)$, which is widely used in the fields of bioinformatics and genome comparison [37]–[40], and in machine learning and cyber security [41]–[43]. It is an information-based measure of the distance between two probability distributions (see Appendix). It serves us as a measure of the human retrospective responsibility for the outcomes of the information analysis stage, $x_a$:

$$Resp(x_a) \overset{\text{def}}{=} \frac{D(x_a, x_{aS})}{D(x_a, x_{aS}) + D(x_a, x_{aH})} \qquad (2)$$

By definition, $Resp(x_a) = 0$ if and only if $x_a = x_{aS}$. In this case, the human self-observed information did not influence the formation of $x_a$. Conversely, if $x_a = x_{aH}$, then $Resp(x_a) = 1$. In this case, the information from the system did not influence the formation of $x_a$, and the human has full responsibility. Lastly, if $x_{aS} = x_{aH}$, $x_a$ will have the same distance from $x_{aS}$ and $x_{aH}$, and hence, $Resp(x_a) = 0.5$. In this case, the information from the intelligent system and the self-observed information contributed equally to the formation of the posterior probability distribution of the environmental states. Other values of $x_{aS}$ and $x_{aH}$ will result in partial human responsibility, ranging between 0 and 1.

The responsibility measure is a relative measure. $D(x_a, x_{aS})$ and $D(x_a, x_{aH})$ can both be close to zero, but $Resp(x_a)$ may still be low, if $D(x_a, x_{aS})$ is much smaller than $D(x_a, x_{aH})$. In this case, as the measure shows, the human's comparative unique contribution to the combined result of the analysis is limited. However, since $D(x_a, x_{aH})$ is small, the human self-observed information alone can provide a reasonably good approximation of $x_a$. To deal with such cases, one should report the actual values in the denominator and numerator in (2), in addition to noting the ratio.

*Retrospective responsibility in action selection*

The measure $Resp(x_a)$ is normative and assumes that the human is rational and combines the self-acquired information and the information from the system optimally. Hence it can be used as a benchmark for the optimal level of retrospective human responsibility for the outcomes of the information analysis stage based on all the available information. However, in reality, people do not always act optimally. In addition, we usually observe only the actual actions the human took and not the subjective formation of the probabilities during the information analysis stage. We, therefore, define a second measure that quantifies how reasonable the specific action the human selected was, given all the information the human had at the decision moment.

We assume that the human selects the action which maximizes benefits and reduces costs. Specifically, we assume that the human selects an action, $x_i$, which maximizes the expected utility, based on $x_a$, which describes the human's posterior probability distribution over all environmental states:

$$\underset{x_s \in X_s}{MAX} U(x_s / x_a) \qquad (3)$$

The maximization result is deterministic and selects the action with the highest expected utility, given the specific information at the decision moment. However, such a non-continuous function of action choice is unrealistic since it implies that even a minuscule utility advantage of one action alternative over the other should always lead to its selection.

Hence, we convert the expected utility values in (3) into action probabilities using the SoftMax (also termed multinomial logistic) function [44], [45], which has become one of the most common random utility models of discrete preferential choices [46]. According to this function, the probability, $p_{x_s}$, of the human to select an action $x_s \in X_s$, is given by:

$$p_{x_s} = \frac{e^{U(x_s / x_a)}}{\sum_{x_r \in X_s} e^{U(x_r / x_a)}} \qquad (4)$$

Equation (4) generates a probability distribution $\{p_{x_i}\}_{x_i \in X_s}$ over all possible human actions, given the optimal use of the available information at the decision moment.

The SoftMax function in (4) takes N real utility values as input, which may contain negative components and might not sum to 1. It normalizes them into a probability distribution consisting of $N$ probabilities, which are proportional to the exponentials of the input values. Larger utility values correspond to larger probabilities, and all probabilities add up to 1. Due to the exponential form of the SoftMax function, it will return a value close to 0 whenever an action alternative value is much less than the maximum value of all the other actions. It will return a value close to 1 when applied to the action with the maximum value unless another action has a similar value. When the absolute difference between the expected utility values of two action alternatives is small, the corresponding probabilities are also close to each other, indicating that it makes sense for the human to choose either action with a similar probability.

From (4), we can generate a probability distribution $\{p_{x_s}\}_{x_s \in X_s}$ of the possibility of selecting each action, given the available information at the decision moment. Denote by $p^*$ the maximum probability, $p^* = Max_{x_s \in X_s}\{p_{x_s}\}$. Assume that the human selected a certain action $x_s$. Then we quantify how reasonable the selection of $x_s$ was:

$$Rsnble(x_s) \overset{\text{def}}{=} \frac{p_{x_s}}{p^*} = \frac{e^{U(x_s / x_a)}}{e^{U(x_s^* / x_a)}} \qquad (5)$$

The ratio of SoftMax probabilities is proportional to the ratio of the exponentials of the corresponding utility values, given all the specific available information at the decision moment. The measure receives a value of 1 if the human selects the optimal action with the largest utility value. It will receive a smaller value, whenever the human chooses a non-optimal action, with a lower utility. However, suppose this action has only slightly lower utility than that of the optimal action. In that case, the ratio will be close to 1, indicating that it is reasonable for the human to choose either of the actions. Finally, if several actions share the same expected utility, their ratio in (5) will be equal, indicating that it is equally reasonable to select either of them.

## III. AN APPLICATION TO BINARY CLASSIFICATION SYSTEMS

We demonstrate the application of the responsibility measures to a DSS, which automatically classifies input into one of two or more categories and may also recommend an action. For simplicity, we focus on binary classification systems that warn users about abnormal conditions. These binary alert systems are widely used on flight decks and vehicles, industrial control rooms, medical equipment and diagnostic procedures, smart homes, and more [47]–[55]. As these systems become more advanced, their users sometimes have to make decisions based exclusively on their output without being able to independently evaluate it, which greatly complicates the determination of human retrospective responsibility.

To calculate the two retrospective measures, we need to make assumptions regarding the nature and interdependencies of the different variables that characterize the environment, the human, and the system.

*Environment states*: Assume that the environment consists of only two types of entities: signals, which should be accepted, and noise, which should be rejected. The relative frequency of signals in the environment is $P_s$, and the relative frequency of noise is $1-P_s$.

*Information acquisition*: Denote by $E$ the set of the observable physical characteristic of the state of the world. The human and the classification system each observe a different uncorrelated measurable property of the state of the world, based on $E$. The distributions of these measurable properties for signal and noise overlap, so there is an ambiguity about whether an observed entity is a signal or noise.

*Information analysis*: Both the human and the classification system try to infer the current environmental state based on the observed properties.

*Action selection:* The classification system classifies events as signals when they are above a preprogrammed threshold and as noise otherwise. The human considers both the human self-acquired information and the classification result of the system when deciding on an action. We assume that the human selects the action that maximizes the expected payoffs.

*Action implementation:* The final outcome (i.e. whether a detected entity was eventually accepted or rejected) is strictly determined by the results of the human action selection process.

Fig. 2 depicts the information flow of human interaction with a binary classification system described above. In this simple case, the four information processing stages can be reduced to analyzing three variables and their inter-dependencies: $e$ (the specific value the human observed), $Y$ (the classification result of the intelligent system), and $X$ (the action selected by the human). These variables dictate the outcome $Z$.

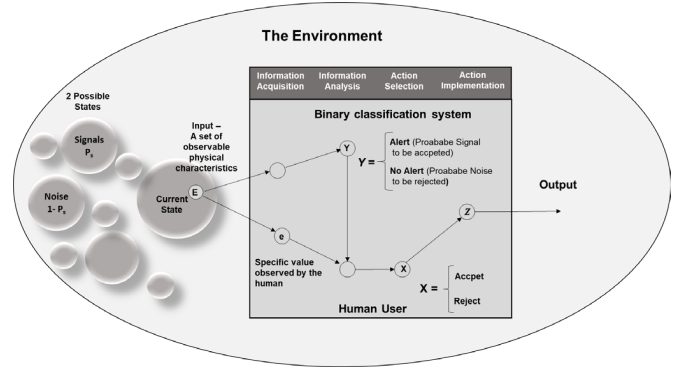

Fig.2. Information flow in the human interaction with a binary classification system.

### A. The unique human share in forming the information used for decision making

When the human has no additional information, a rational human's estimate of the probability for signal and noise would be their known relative frequency in the environment, $P_s$, and $1-P_s$, respectively.

Assume that the human knows the performance properties of the classification system, namely its expected rates of True Positive, False Negative, False Positive, and True Negative, which will be denoted, respectively, by $\hat{P}_{TP}$, $\hat{P}_{FN}$, $\hat{P}_{FP}$, and $\hat{P}_{TN}$. Using Bayes' law, the human can use these rates to update the prior probabilities for signal and noise based on the system's classification results. In this case, $x_{aS}$, the posterior probability distribution for signal and noise, when the human uses only information presented by the classification system, is given by:

$$x_{aS} = \left( \frac{P_s \hat{P}_{TP}}{P_s \hat{P}_{TP} + (1-P_s) \hat{P}_{FP}}, \frac{(1-P_s) \hat{P}_{FP}}{P_s \hat{P}_{TP} + (1-P_s) \hat{P}_{FP}} \right) \quad (6a)$$

when the system classified an entity as signal, and

$$x_{aS} = \left( \frac{P_s \hat{P}_{FN}}{P_s \hat{P}_{FN} + (1-P_s) \hat{P}_{TN}}, \frac{(1-P_s) \hat{P}_{TN}}{P_s \hat{P}_{FN} + (1-P_s) \hat{P}_{TN}} \right) \quad (6b)$$

when the system classified an entity as noise.

Assume that the human self-acquired a specific value $e$ for the measurable property of the state of the world. In this case, $x_{aH}$, the posterior probability distribution for signal and noise when the human uses only self-acquired information is:

$$x_{aH} = \left( \frac{P_s f_s(e)}{P_s f_s(e) + (1-P_s) f_N(e)}, \frac{(1-P_s) f_N(e)}{P_s f_s(e) + (1-P_s) f_N(e)} \right) \quad (7)$$

where $f_s(e)$ and $f_N(e)$ are, respectively, the probability density functions of signal and noise.

By combining both sources of information, the human can compute $x_a$, posterior probability distribution for signal and noise, using all the information known to the human.

$$x_a = \left( \frac{P_s \hat{P}_{TP} f_s(e)}{P_s \hat{P}_{TP} f_s(e) + (1-P_s) \hat{P}_{FP} f_N(e)}, \frac{(1-P_s) \hat{P}_{FP} f_N(e)}{P_s \hat{P}_{TP} f_s(e) + (1-P_s) \hat{P}_{FP} f_N(e)} \right) \quad (8a)$$

when the system classified an entity as a signal and the human observed $e$, and

$$x_a = \left(\frac{P_S \bar{P}_{FN} f_S(e)}{P_S \bar{P}_{FN}(e) + (1 - P_S)\bar{P}_{TN} f_N(e)}, \frac{(1 - P_S)\bar{P}_{TN} f_N(e)}{P_S \bar{P}_{FN}(e) + (1 - P_S)\bar{P}_{TN} f_N(e)}\right) \quad (8b)$$

when the system classified an entity as a noise and the human observed $e$.

The computation of $Resp(x_a)$, the retrospective human responsibility for the outcomes of the information analysis, is straightforward from equations (6), (7), and (8).

### B. The reasonability of the selected action

Denote by $V_{TP}$, $V_{FN}$, $V_{FP}$, $V_{TN}$ the payoffs the human associates with correct and incorrect acceptance and rejection of signal and noise (in this form of presentation, $V_{FP}$ and $V_{FN}$ are negative). Denote by $\hat{P}_s$ and by $1 - \hat{P}_s$, respectively, the estimation of the probabilities of signal and noise based on all available information at the decision moment. Then, $x_a = (\hat{P}_s, 1 - \hat{P}_s)$, is given by either (8a) or (8b), depending on the specific result of the classification system. The corresponding expected values are:

$$EV_{Accept} = \hat{P}_s \cdot V_{TP} + (1 - \hat{P}_s) \cdot V_{FP} \quad (9)$$

$$EV_{Reject} = \hat{P}_s \cdot V_{FN} + (1 - \hat{P}_s) \cdot V_{TN} \quad (10)$$

According to the multinomial logit model, the probability that the human should prefer an action alternative is given by:

$$P_{Accept} = 1 - P_{Reject} = \frac{e^{EV_{Accept}}}{e^{EV_{Accept}} + e^{EV_{Reject}}} \quad (11)$$

Hence, the reasonability for selecting each action is given by:

$$Rsnble("Accept") = \frac{P_{Accept}}{Max(P_{Accept}, P_{Reject})}$$

$$= \frac{e^{EV_{Accept}}}{Max(e^{EV_{Accept}}, e^{EV_{Reject}})}$$

$$\quad (12)$$

$$Rsnble("Reject") = \frac{P_{Reject}}{Max(P_{Accept}, P_{Reject})}$$

$$= \frac{e^{EV_{Reject}}}{Max(e^{EV_{Accept}}, e^{EV_{Reject}})}$$

## IV. DEMONSTRATIONS OF RETROSPECTIVE RESPONSIBILITY COMPUTATIONS

### A. SDT based model

Computations require assumptions regarding the probabilistic distributions and interdependencies of the different variables that characterize the environment ($e$), the human ($X$), and the system ($Y$). We employ the assumptions and formulation of Signal Detection Theory (SDT) [56], [57]. This well-established approach is widely applied in fields such as psychology, decision-making, machine learning (statistical classification), alarm management, medical diagnostics, security, and more.

According to the equal variance Gaussian SDT model, the distributions of signal and noise characteristics are Gaussian with equal variance and different means, which allows some discrimination between the two. However, the distributions overlap, so there is some remaining uncertainty.

When describing a detector, SDT differentiates between its detection sensitivity and response criterion. The detection sensitivity ($d'$) is the detector's ability to distinguish between signal and noise. It is defined as the distance between the means of the signal and noise distributions, measured in standard deviations. When $d'=0$, the detector is unable to distinguish between signal and noise. The larger the detection sensitivity, the better the detector's distinguishing ability. We will denote by $d'_{System}$ and $d'_{Human}$, respectively, the detection sensitivities of the binary classification system and the human.

The response criterion ($\beta$) represents the detector's tendency to favor one response over the other, which is determined by a utility function that relates costs and benefits to different outcomes. We assume the binary classification system has a preset fixed response criterion, denoted by $\beta_{System}$, which determines its output. According to SDT, when a human works alone, without the aid of a system, the optimal human response criterion, $\beta_{Human}$, that maximizes the expected payoffs, is:

$$\beta_H^* = \frac{1 - p_s}{p_s} \frac{V_{TN} - V_{FP}}{V_{TP} - V_{FN}} \quad (13)$$

When aided by the classification system, the human can use its output as additional input to self-acquired information, considering the system's expected rates of true positive and false positive output. In this case, the optimal human behavior is to use two different response criteria, depending on the system's classification results. These criteria are computed by replacing $P_S$ in (13), with $P_{S|A}$ - the conditional probability for a signal, given that the system initiated an alert for signal, or $P_{S|NA}$ - the conditional probability for a signal, given that the system did not issue such alert. When using a reliable classification system, the posterior probability for a signal is larger when the system indicates a signal and lower when the system indicates noise, $P_{S|A} \geq P_S \geq P_{S|NA}$. Hence, the human should increase the tendency to accept the entity as a signal when the system alerted for signal, and increase the tendency to reject the entity as noise when the system did not indicate a signal.

### B. Numerical example

*Example*: Assume a binary classification system with a somewhat better detection sensitivity than that of the human: $d'_{System} = 2.0$ and $d'_{Human} = 1.5$. Assume a prior signal frequency of $P_t = 0.2$ and a payoff scheme of $V_{TP} = 10, V_{TN} = 10, V_{FP} = -10, V_{FN} = 20$, which are the expected values for different outcomes. This payment scheme relates higher costs to missing a signal. Assume that both the system and the human have similar incentives and use a response criterion that matches the above payoff scheme and the signal frequency. The corresponding rates of the classification system are $\tilde{P}_{TP} = 0.69$, $\tilde{P}_{FN} = 0.31$, $\tilde{P}_{FP} = 0.07$, and $\tilde{P}_{TN} = 0.93$.

Assume now that the human observed a value of –1.5, the classification system classified the entity as a "Signal", but the human chose to ignore this indication and rejected the entity as "Noise". Eventually, the entity was indeed a signal, as the system pointed out, which led to a loss of -20. In this setting, what was the human's comparative contribution in forming the information for decision-making? How reasonable was the human decision?

*Answer:* From 6(a), when the human relies only on the information from the system, which classified the entity as a signal, the posterior probability distribution of the environmental states is:

$$x_{aS} = \left( \frac{0.2 \cdot 0.69}{0.2 \cdot 0.69 + 0.8 \cdot 0.07}, \frac{0.8 \cdot 0.07}{20.69 + 0.8 \cdot 0.07} \right) = (0.72, 0.28) \quad (14)$$

Hence, based only on the information from the system, the posterior probability for a signal is 72%. However, the value the human observed (-1.5) contradicts this conclusion as it strongly indicates noise. In this case, $f_S(-1.5) = 0.004$ and $f_N(-1.5) = 0.13$, so from (7) we have:

$$x_{aH} = \left( \frac{0.2 \cdot 0.004}{0.2 \cdot 0.004 + 0.8 \cdot 0.13}, \frac{0.8 \cdot 0.13}{0.2 \cdot 0.004 + 0.8 \cdot 0.13} \right) = (0.01, 0.99) \quad (15)$$

If the human relies only on self-acquired information, the probability of a signal is only 1%.

When optimally combining both types of information, taking into consideration the differences between the human and system detection sensitivities, we have from (8a) a signal probability of 8%:

$$x_a = \left( \frac{0.2 \cdot 0.69 \cdot 0.004}{0.2 \cdot 0.69 \cdot 0.004 + 0.8 \cdot 0.07 \cdot 0.13}, \frac{0.8 \cdot 0.07 \cdot 0.13}{0.2 \cdot 0.69 \cdot 0.004 + 0.8 \cdot 0.07 \cdot 0.13} \right) =$$
$$= (0.08, 0.92) \quad (16)$$

Hence, the human indication for noise is so strong that the contradicting information from the classification system has only a limited effect on the combined posterior probabilities. This is reflected in the Jensen-Shannon distance between the above three posterior probability estimations for signal and noise. The human responsibility for the outcomes of the information analysis stage is:

$$Resp(x_a) \overset{\text{def}}{=} \frac{D(x_a, x_{aS})}{D(x_a, x_{aS}) + D(x_a, x_{aH})} = \frac{0.58}{0.58 + 0.16} = 79\% \quad (17)$$

Hence, in the above settings, if the human optimally combines the information from both sources, the human provides most of the information used for the decision, even though the human has a lower detection sensitivity.

Based on $x_a = (0.08, 0.92)$, the corresponding expected values are:

$$EV_{Accept} = 0.08 \cdot 10 + 0.92 \cdot (-10) = \text{-8.4} \quad (18)$$

$$EV_{Reject} = 0.08 \cdot (-20) + 0.92 \cdot (10) = \text{7.6} \quad (19)$$

Thus,

$$Rsnble("Accept") = \frac{e^{EV_{Accept}}}{Max(e^{EV_{Accept}}, \ e^{EV_{Reject}})} \approx 0 \quad (20)$$

$$Rsnble("Reject") = \frac{e^{EV_{Reject}}}{Max(e^{EV_{Accept}}, \ e^{EV_{Reject}})} = 1$$

The human decision to reject was entirely reasonable, even if it eventually led to adverse outcomes, which in this case are expected 8% of the time.

The above example used specific values of $d'_{System}$ and $d'_{Human}$. In the next sections, we analyze the retrospective measures as a function of $d'_{System}$ and $d'_{Human}$, each on a scale ranging between .6 (low ability to distinguish between signal and noise) and 3 (high ability to distinguish between signal and noise). We assumed that the intelligent system classified the entity as "signal" in all cases. We used the same prior signal frequency ($P_t = 0.2$) and payoff scheme ($V_{TP} = 10, V_{TN} = 10, V_{FP} = -10, V_{FN} = -20$) as in the above example.

### C. The unique human share in forming the information used for decision making

Fig. 3 presents the comparative unique human share as a function of $d'_{System}$ and $d'_{Human}$ for different specific values for the human self-acquired information, $e$, ranging between -1.5 (strong indication for noise) and 4.5 (strong indication for signal). We assumed that, in all cases, the intelligent system classified the entity as "signal".

Fig.3 (a) presents a case where the human observed $e = -1.5$, which is a very strong indication of noise that contradicts the information for signal from the classification system. In this case, based on the human self-acquired information alone, the probabilities for a signal are in the range of 0%-8%. The indication for noise is so strong that the contradicting information from the classification system has only a limited effect on the overall combined posterior probabilities for signal and noise. Thus, in most combinations of human and system detection sensitivities, the comparative contribution of human-based information to the outcomes of the information analysis is high (80%-100%). Only when the detection sensitivity of the human is low, and the system's detection sensitivity is high should a rational human give less weight to the self-acquired information, leading to lower comparative human contribution. As one may predict, the comparative contribution of the human self-acquired information increases in $d'_{Human}$ and decreases in $d'_{System}$.

Fig.3 (b) presents a case where the human observed $e = 0$. In this case, based on the human self-acquired information alone, the probabilities for a signal are somewhat higher but still quite low (0%-17%). Since the human self-acquired information is less conclusive, there are more combinations in which the human should give more weight to the information from the classification system. Again, the comparative contribution of the human self-acquired information to the outcomes of the information analysis increases in $d'_{Human}$ and decreases in $d'_{System}$.

Fig.3 (c) presents cases where the human observed $e = 1.5$. In this case, based on the human self-acquired information alone, the probabilities for signal and noise are of similar magnitude, so the human is less conclusive about the entity type. Hence, the human gives more weight to the information from the classification system. This weight increases as the classification system is more accurate, so the human comparative contribution decreases with $d'_{System}$. However, because of the Gaussian shape of the density function of signal and noise, the human's comparative contribution is not monotonous in $d'_{Human}$. As $d'_{Human}$ increases, the likelihood ratio between signal and noise increases until a maximum at $d'_{Human}=e$, and then it decreases. Hence, for a given fixed value of $d'_{System}$, the combined signal probability also increases to a maximum at $d'_{Human}=e$ and then decreases.

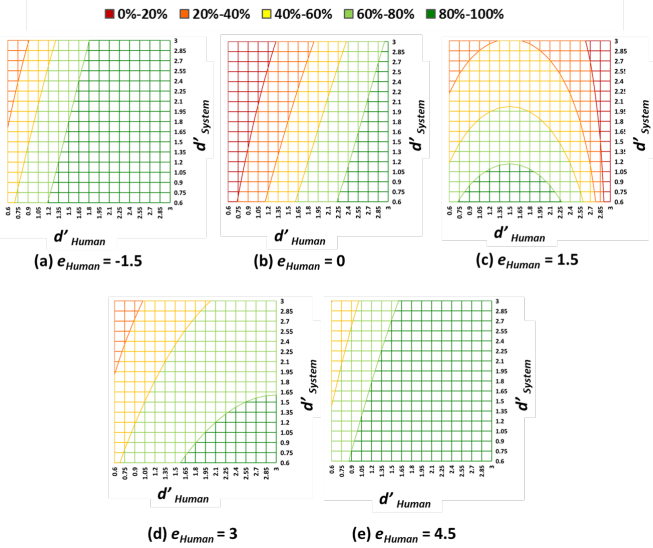

Fig.3. The unique human contribution to the information used for decision-making, $Resp(x_a)$, when the classification system classified the entity as a signal for different values of human and system detection sensitivities and human observed values.

In the cases presented in Fig.3 (d), (e), the human observed a strong indication for signal, which matches the information from the classification system. Hence, when combining the information from the two, the posterior probability for a signal increases. In the examined range of detection sensitives, the comparative contribution of the human self-acquired information increases in $d'_{Human}$ and decreases in $d'_{System}$.

To conclude, the above numerical results have intuitive explanations. The unique human share in forming the information used for decision-making depends on the differences between the human and system detection sensitivities and the uncertainty reduction due to human self-acquired information. The human should give more weight to the information from the classification system when the system has known superior detection capabilities, so the comparative unique human contribution decreases with $d'_{System}$. However, when the human has substantial self-acquired information, the results of the classification have a lesser effect, leading to a higher comparative human contribution.

### D. The reasonability of the selected action

Figure 4 shows the reasonability of the human's decision to accept or reject.

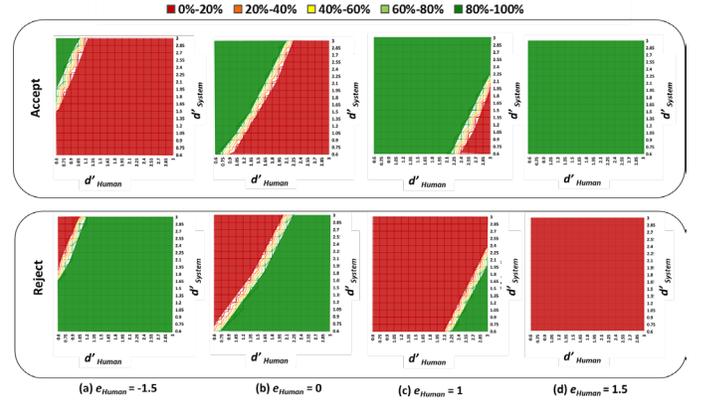

Fig.4. The reasonability to select different actions, $Rsnble(x_s)$, when the classification system classified the entity as a signal, for different values of human and system detection sensitivities and observed values by the human.

Since the classification system indicated a signal, the combined probability for signal increases when the human self-acquired information indicates a higher probability for signal, making it more reasonable to accept as $e$ increases. When the human self-acquired information contradicts the information from the classification system, the value of $Rsnble(x_s)$ depends on the specific combination of human and system detection sensitivities and on the uncertainty reduction, due to the human self-acquired information.

In most combinations of human and system detection sensitivities, the selection of a certain action is fully or not at all justified. There are only narrow margins in which the Reasonability of an action will not be close to 0 or 1. This is because we transformed the expected utility values into action selection probabilities, using the SoftMax function. With this exponential transformation, the action alternative with a higher excepted value usually has a probability close to 1, while the other action alternative has a probability close to 0. Hence their corresponding $Rsnble(x_s)$ values are also almost binary.

The SoftMax probabilities will only be close to each other when the two action alternatives have very similar expected values. In this case, the alternative with the higher expected value will have a reasonability measure of 1. In contrast, the other alternative will receive a value close to 1, indicating that it is reasonable for the human to choose it also. For example, in Fig. 4(b), such is the case at the point $d'_{Human} = 1.35$ $d'_{System} = 1.5$. At this point, the expected utility values are close, so the corresponding SoftMax probabilities are PAccept = 0.56 and $P_{Reject}$ = 0.44. Hence, in this case $Rsnble(Accept)$ = 100%, and $Rsnble(Reject)$ = 0.44/0.56 =78%.

To conclude, the above numerical analysis has intuitive explanations. When the human self-acquired information presents a strong indication for either signal or noise, the results of the classification system should not affect the human decision. In this case, it is more reasonable for humans to act according to self-acquired information. When the classification system is much more accurate than the human or self-acquired information is less indicative, a rational human should follow the classification system recommendation. Selecting the alternative with the higher expected value (Rsnble(X)= 100%) is always fully reasonable. The other alternative has a value between 0 (when its expected value is considerably lower) or 100% (when the two alternatives have the same expected value).

## V. Conclusion and implications

We extended the ResQu model [33] to include *retrospective* responsibility measures that quantify human responsibility for past events that involved intelligent systems. When the human interacts repeatedly with a system, one can infer the underlying distributions from empirical observations taken over time and compute a responsibility measure, which quantifies the *average* unique share of the human in determining the output distribution.

However, this measure cannot be applied for the analysis of a *single* past event. In that case, we introduced two additional measures. The first quantifies the unique human share in forming the information used for decision-making. The second quantifies how reasonable the specific human action was based on all the available information at the time of the decision.

The retrospective responsibility measures are influenced by the combined convoluted effects of system design, the human role and authority, and probabilistic factors related to the system and the environment. Therefore, even when human operators fulfill essential system functions and roles, they may still not be fully responsible for adverse outcomes. Hence, there could be situations where the human would be considered responsible for adverse outcomes, even when contributing little to the outcomes. In this case, humans may be the ones bearing moral and legal responsibility for system faults, acting as "moral crumple zones" [58], [59]. Our retrospective measures can help to resolve this case by providing a tool for quantifying different aspects of the unique human contribution.

For example, in the above-mentioned 2011 UBS unauthorized trading case, we can consider the anti-fraud system's specific expected rates of true positive and false positive indications and the independent ability of employees to detect fraud without the system. We can compute the employee's unique contribution to the decisions based on these. By associating human utility values with correct and incorrect actions, we can also evaluate how reasonable the decision to ignore the alert from the anti-fraud system was.

To conclude, we developed retrospective measures to quantify different aspects of human causal responsibility in past interactions with intelligent systems. These measures can serve as an additional tool in investigating and analyzing past events, which require the judgment of human responsibility for outcomes. They provide a novel quantitative perspective and can also aid in policy and legal decisions.

Future work should expand the model, enabling it to deal with temporal effects, such as the time needed to make a decision and its effects on the human's tendency to depend on the intelligent system and comply with its decisions and actions. The information-theoretical framework should be expanded to deal with temporal aspects by evaluating both transmitted information and information transmission rates considering human channel capacity constraints.

## VI. Appendix – The Jensen–Shannon distance

In information theory, the Jensen–Shannon divergence (JSD) measures the similarity between two probability distributions. It is a smoothed version of the Kullback–Leibler divergence (KLD), with some notable differences, mainly that it is symmetric and always has a finite value. The JSD between two probability distributions, $P$ and $Q$, is defined as:

$$JSD(P \parallel Q) \overset{\text{def}}{=} \tfrac{1}{2}KLD(P \parallel M) + \tfrac{1}{2}KLD(Q \parallel M) \quad (A1)$$

Where $M$ is the arithmetic mean of the two probability distributions, $M = \tfrac{1}{2}(P + Q)$, and KLD is the Kullback–Leibler divergence, which describes the amount of information lost when a probability distribution $S$ is used to approximate a reference probability distribution $R$,

$$KLD(P \parallel Q) \overset{\text{def}}{=} \sum_{x \in \chi} R(x) \cdot [log_2 R(x) - log_2 S(x)] =$$

$$= \sum_{x \in \chi} R(x) \cdot log_2\left(\frac{R(x)}{S(x)}\right) \quad (A2)$$

Hence, we can write (2) explicitly as:

$$JSD(P \parallel Q) \overset{\text{def}}{=} \ \tfrac{1}{2}\sum_{x \in \chi} P(x) \cdot log_2\left(\frac{P(x)}{M(x)}\right) +$$

$$+ \tfrac{1}{2}\sum_{x \in \chi} Q(x) log_2\left(\frac{Q(x)}{M(x)}\right) \quad (A3)$$

where $M = \tfrac{1}{2}(P + Q)$. The square root of the JSD is a metric, often referred to as Jensen-Shannon distance [60], [61].

$$D(P,Q) \overset{\text{def}}{=} \ \sqrt{JSD(P \parallel Q)} \quad (A4)$$

The Jensen-Shannon distance, $D(P,Q)$, is a normalized distance, which is symmetric, satisfies the triangle inequality, and is bounded between 0 and 1. It equals 0 iff $P=Q$ almost everywhere, and it reaches its maximum value of 1 when the supports of the two distributions do not intersect each other.

The $log_2$ function in (A3) is undefined at Zero. However, if there is an $x$, such that both $P(x) = 0$ and $Q(x)=0$, then $M(x)=0$ as well. In this case, we define $log_2\left(\frac{P(x)}{M(x)}\right) = log_2\left(\frac{Q(x)}{M(x)}\right) = 0$, so there is no contribution from $x$ to the summation in the Jensen–Shannon distance. In addition, if there is an $x$, such that $P(x) = $ 0 but $Q(x)$=q > 0, then $M(x)$ = q/2 > 0. In this case $log_2\left(\frac{P(x)}{M(x)}\right)$ = $log_2(0)$ is undefined but $\lim_{P(x)\to 0^*} P(x) * log_2\left(\frac{P(x)}{M(x)}\right) = 0$, so $x$ will have no contribution to $KLD(P||M)$ and a positive contribution to $KLD(Q||M)$.